\documentclass[a4paper]{jpconf}
\usepackage{graphicx}
\begin{document}
\title{Optical followup of galaxy clusters detected by the South Pole Telescope}

\author{S.Desai$^1$, R.Armstrong$^2$, M.L.N.Ashby$^3$, B.Bayliss$^3$, G.Bazin$^1$, B.Benson$^4$, E.Bertin$^5$, L.Bleem$^4$, M.Brodwin$^6$, A.Clochiatti$^7$, R.Foley$^3$, M.Gladders$^4$, A.H.Gonzalez$^8$, F.W.High$^4$, J.Liu$^1$, J.Mohr$^1$, A.Rest$^9$, J.Ruel$^{10}$, A.Saro$^1$, J.Song$^{11}$, B.Stalder$^3$, A.Stanford$^{12}$, C.Stubbs$^{10}$, A.Zenteno$^1$}

\address{$^1$ Dept. of Physics, Ludwig-Maximilians-Universit\"{a}t, Scheinerstr.\ 1, 81679 M\"{u}nchen, Germany}
\address{$^2$ Dept. of Physics and Astronomy, University of Pennsylvania, Philadelphia, PA 19104, USA}
\address{$^3$ Harvard-Smithsonian Center for Astrophysics, 60 Garden Street, Cambridge, MA 02138, USA}
\address{$^4$ Dept of Astronomy and Astrophysics, University of Chicago, 5640 South Ellis Avenue, Chicago, IL 60637, USA}
\address{$^5$ Institut d'Astrophysique de Paris, 98 bis boulevard F-75014 Parago, Paris, France}
\address{$^6$ Dept. of Physics, University of Missouri, 5100 Rockhill Road, Kansas City, MO 64110, USA}
\address{$^7$ Dept. de Astronomia y Astrofisica, Ponticia Universidad Catolica de Chile, Casilla 306, Santiago 22, Chile}
\address{$^8$ Dept. of Astronomy, University of Florida, Gainesville, FL 32611, USA}
\address{$^9$ Space Telescope Science Institute, 3700 San Martin Dr., Baltimore, MD 21218, USA}
\address{$^{10}$ Dept. of Physics, Harvard University, 17 Oxford Street, Cambridge, MA 02138, USA}
\address{$^{11}$ Dept. of Physics, University of Michigan, 450 Church St. Ann Arbor, MI 48109, USA}
\address{$^{12}$ Dept. of Physics, University of California, 1 Shields Ave. Davis, CA 95616, USA}

\ead{shantanu@usm.lmu.de}

\begin{abstract} The South Pole Telescope (SPT) is a 10 meter telescope operating at mm wavelengths. It  has
recently completed a three-band survey covering 2500 sq. degrees. One of the survey's main goals  is to
 detect galaxy clusters using Sunyaev-Zeldovich effect and use these clusters for a variety of
 cosmological and astrophysical studies such as the dark energy equation of state, the primordial 
non-gaussianity and the evolution of galaxy populations. Since 2005, we  have been engaged in  a
 comprehensive optical and 
near-infrared followup program (at wavelengths between 0.4 and 5 $\mu$m) to image high-significance SPT clusters,
 to measure their photometric redshifts, and  to estimate the contamination rate of the candidate lists. 
These clusters are then used for various cosmological and astrophysical studies.

\end{abstract}

\section{Introduction}

The abundance and distribution of galaxy clusters are a powerful probe of the dark energy equation of state and non-gaussianity, 
since they depend upon both the expansion history of the universe and  the growth of density
 fluctuations~[1].  The Sunyaev-Zel'dovich Effect (SZE)~[2] is one
 of the most promising methods of discovering galaxy clusters and measuring this abundance as a function of 
cosmic time. The SZE is a  spectral distortion of the microwave background spectrum caused
 by  the inverse Compton scattering of CMB photons by hot cluster electrons.  
The surface brightness of the  SZE is nearly redshift independent, enabling SZE surveys to detect clusters out 
to the earliest epochs at which they exist. Thus, SZE selected samples of galaxy clusters enable powerful tests of structure formation models and of the cosmic acceleration, but only when the
 cluster redshifts and masses  are accurately known. Although  the cluster mass can be estimated from the 
SZE detection significance, the SZE signal alone does not provide an estimate of cluster redshift.
 These  redshifts can only be obtained via followup observations at optical and near-infrared wavelengths.

\section{South Pole Telescope}
The South Pole Telescope (SPT) is a 10 meter telescope with  a $1^{\circ}$ field of view,
taking data at mm wavelengths, that has just  completed a deep three-band (95, 150 and 220 GHz) 
arc-minute resolution survey covering 2500 sq. degrees of  the southern extragalactic sky ($20~hr<\alpha<24~hr$, $0~hr<\alpha<7~hr$ and
 $-65^{\circ} < \delta < -40^{\circ}$). One of the primary goals of SPT is to detect galaxy clusters using the SZE effect. First  observations of SZE-selected galaxy clusters from SPT were reported in 2008 [3]. Potential galaxy cluster candidates along with their detection significance are obtained from SPT observations using a matched filtering technique and the detection significance is well correlated with the cluster mass [4].
 We make use of followup optical observations using the positions of SPT cluster candidates to confirm
galaxy clusters and measure their redshifts.

\section{Optical and near-infrared followups}

\subsection{DSS vetting}
To optimize the use of available telescope resources and time, we use Digitized Sky Survey (DSS) images
to  make a rough segregation of our cluster candidates into high and low redshift samples.
 We find that most clusters at $z < 0.5$ are visible in these photographic plates, with a  tail extending to
 redshift of $\sim$ 0.7. We inspected for evidence of massive clusters in the DSS plates and assigned scores from zero to three, where zero implies no evidence for a cluster and three implies extensive evidence for a cluster. The contamination rate for DSS scores greater than zero is only a few  percent.  Roughly 50\% of SPT cluster candidates have DSS scores greater than zero.


\subsection{Photometric optical followups}

The main goal of optical observations of galaxy clusters is to achieve enough depth to sample the red
sequence of a cluster, down to $0.5L^{*}$ with
5 $\sigma$ photometry. We have been doing two kinds of 
follow-ups : survey-mode where we uniformly cover a contiguous area and pointed follow-up. 
The first contiguous followup program
was the Blanco Cosmology Survey (BCS) which was a 4-band $griz$ survey from 2005-2008 in
two 50 sq. degree patches centered on RA of 5 hr and 23 hrs and declination of $\sim -55^{\circ}$.
We expect to cover the entire SPT region once the Dark Energy Survey begins in 2012.

The other kind of followup program is using pointed observations of SPT cluster candidates with SNR 
greater than 4.5. All clusters not seen in DSS  (therefore $z > 0.5$ or contamination) are followed up with 
4m-6.5m  class telescopes, whereas those seen in DSS are followed up with telescopes with smaller apertures, including the
 SWOPE 0.9m  and ESO/MPG 2.2m.  
This program has been ongoing since 2008 and a summary of different  telescopes and
the number of nights observed can be found in Tab.~\ref{tab:followup}.  For the bigger telescopes, we have been using an adaptive strategy. We first image the potential cluster candidate in $grz$ bands and if it is  confirmed with $z > 0.7$, we take further observations in $i$ band. 
As of Sept. 2011, 597 clusters above SNR of 4.5 have been observed. From a study of a complete sample of 226 candidates covering
the first 750 sq. degrees, the mean redshift is about 0.55 and  purity is about 80\% (95\%) 
for $S/N > 4.5 (5)$. A full  20\% of the sample is at $z > 0.8$.

\subsection{Near-infrared and spectroscopic followups}
Clusters at high redshifts ($z > 1$) are best observed in near-infrared (NIR).  If the cluster is not confirmed after $griz$ photometry, 
then it is scheduled for NIR observations. For this purpose we use the IRAC on SPIZER
telescope and the NEWFIRM infrared imager on the Blanco 4m telescope. 
For a significant fraction of  the SPT candidates, we have short-exposure Spitzer  imaging at 3.6 $\mu$m and 4.5 $\mu$m to confirm clusters
that extend to z  $\sim$ 1.35. We try to observe all high-redshift clusters with $z > 1$ in NEWFIRM $J$ and $K$ bands and unconfirmed clusters in $K$ bands. As of Sept. 2011, 296 clusters have been followed up on Spitzer and 52 by NEWFIRM.

We have also received Gemini time for a spectroscopic study of 11  X-ray (Chandra and XMM) targeted clusters, and
 we have recently been awarded a GMOS spectroscopic survey program to follow up a sample of 80 clusters 
in the redshift range 0.5 to 0.8. About 17  of those systems have been observed so far, and 44 more are still targeted for observations. We have observed 25 clusters with VLT/FORS2 at $z > 0.8$ for dynamical mass calibration.  We are also doing weak-lensing mass 
measurements using HST+VLT at high redshifts,  and Magellan out to $z \sim 0.6$.

\subsection{Photometric redshift estimation}
The data from optical imaging observations are processed and calibrated using two independent pipelines which have been developed for Dark Energy Survey and Super-MACHO projects.  The calibrated colors are then used to estimate photometric redshifts.
The main principle used to obtain photometric redshifts is to fit the observed colors of the galaxy populations of 
potential cluster  candidates to red-sequence   population synthesis models~[6] near the SZE cluster coordinate and to use the redshift 
which maximizes the signal over background [7].  A comparison of photometric and
 spectroscopic redshifts (as of Sept. 2011) is provided in Fig.~\ref{photoz}. The photo-z accuracy is,   $\sigma_z/(1+z) \sim 2-3\%$. 

\section{Selected science results}
\begin{figure}
\begin{minipage}{16pc}
\includegraphics[width=16pc]{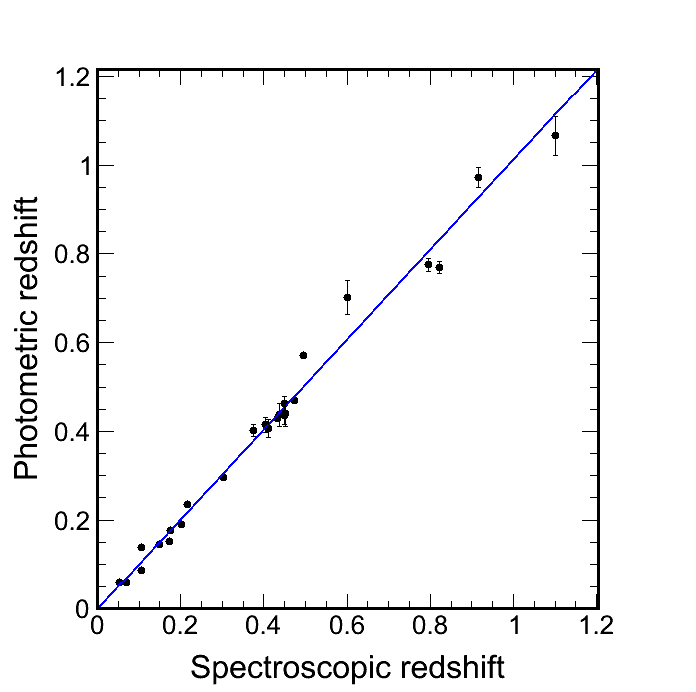}
\caption{Comparison of photometric and spectroscopic redshifts of SPT galaxy clusters. The characteristic scatter is $\sigma_z/(1+z) \sim 2-3\%$.}
\label{photoz}
\end{minipage}
\hspace{0.4cm}
\begin{minipage}{18pc}
\includegraphics[width=18pc]{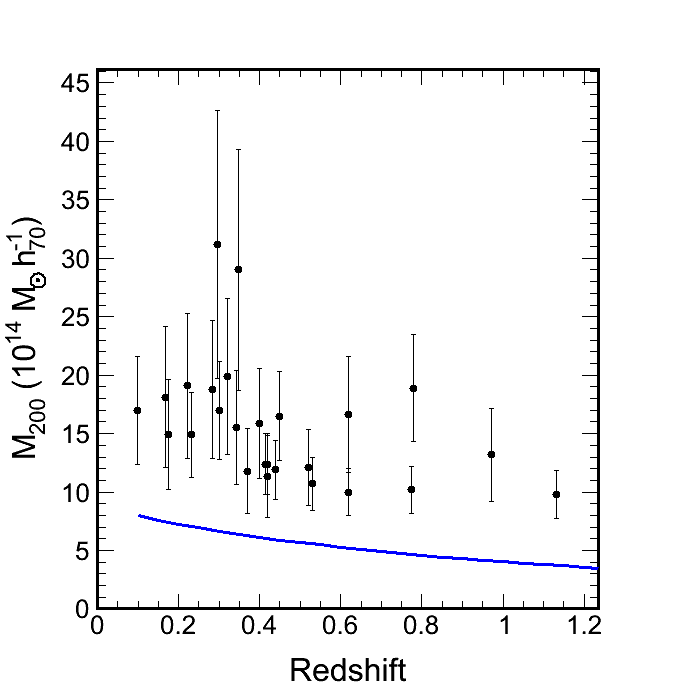}
\caption{Redshift-mass distribution for the 26 most massive SPT clusters from 2500 sq. degree survey [5].
The solid line marks the mass limit for the full sample   down to signal to noise of 4.5, where the redshift followup 
is not yet complete.}
\label{mass}
\end{minipage}
\end{figure}

We have done a variety of cosmological and astrophysical studies based on optical followups of SPT 
galaxy clusters (Fig.~\ref{mass}) and several projects are currently underway. 
We have reported photometric redshifts and richness estimates from first 178 sq. degrees of SPT observations using BCS and pointed optical observations [7] along with cosmological 
analysis of 21 clusters [8]. We presented multi-wavelength observations of  two massive 
clusters SPT-CLJ0546-5345 ($z=1.067$) [9] and  SPT-CLJ2106-5844 ($z=1.13$) [10] along with 
their implications for  $\Lambda$CDM cosmology. 
We discussed first constraints on non-Gaussianity using 26 most significant  SZE clusters 
seen in SPT survey~[5]. We also reported on  various studies of galaxy populations in the first four SPT clusters using BCS data [11] as well as joint SPT/optical observations of some Planck clusters [12].
Currently we are completing the analysis   of a complete sample covering 750 sq. degrees 
while continuing to pursue   the full redshift followup.
\medskip

\begin{center}
\begin{table}[h]
\centering
\caption{\label{tab:followup} Summary of ground-based optical/NIR photometric followups of SPT galaxy clusters.} 
\begin{tabular}{@{}l*{15}{l}}
\br
Telescope/Imager & Aperture & Nights&FOV ($\circ$) &Footprint ($\circ^2$) \\
\mr
Swope & 0.9m & 25 & 0.07 & 18 \\
Blanco/Mosaic-II & 4m & 45 & 0.68 & 93 \\
Magellan/LDSS3 & 6.5m & 18 & 0.018 & 4 \\ 
Magellan/IMACS & 6.5m & 23 & 0.22 & 22 \\
ESO/MPG/WFI  & 2.2m & 5 & 0.5 & 0.4 \\
NTT & 3.5m & 3 & 0.5 & 0.5 \\
SOAR/SOI & 4m & 5 & 0.09 &  1.7 \\
Blanco/Mosaic-II (BCS) & 4m & 60 & 0.68 &  80 \\
Blanco/NEWFIRM & 4m & 5 & 0.47 & 2.5 \\
\br
\end{tabular}
\end{table}
\end{center}

\ack 
The SPT is supported by the National Science Foundation through grants ANT-0638937 and ANT-0130612.

\end{document}